# An Algorithm to Determine Energy-aware Maximal Leaf Nodes Data Gathering Tree for Wireless Sensor Networks

Natarajan Meghanathan
Assistant Professor of Computer Science
Jackson State University
Jackson, MS 39217, USA
Phone: 601-979-3661; Fax: 601-979-2478
E-mail: natarajan.meghanathan@jsums.edu

**Abstract**

We propose an Energy-aware Maximal Leaf Nodes Data Gathering (EMLN-DG) algorithm for periodic data collection and transmission in wireless sensor networks. For each round of data gathering, an EMLN-DG tree spanning the entire sensor network is formed based on the residual energy level available at the nodes and the number of uncovered neighbors of a node during tree formation. Only nodes that have a relatively larger number of neighbors as well as a higher energy level are included as intermediate nodes in the EMLN-DG tree. By maximizing the number of leaf nodes in a DG tree and considering the energy level available at the nodes while forming the tree, we reduce energy consumption per round as well as balance the energy level across all the nodes in the network. This contributes to a significantly larger network lifetime, measured as the number of rounds before the first node failure due to exhaustion of battery charge. Performance comparison studies with the well-known data gathering algorithms such as LEACH and PEGASIS illustrate that EMLN-DG can help to sustain the network for a significantly larger number of rounds and at the same time incur a lower, or if not comparable, energy loss, delay and energy loss*delay per round of data gathering.

**Key Words:** Data Gathering Trees, Sensor Networks, Maximal Leaf Nodes, Energy-awareness, Network Lifetime, Data Aggregation, Algorithm

## 1 Introduction

A wireless sensor network is often viewed as a distributed system of smart sensor nodes that gather data about the ambient environment and propagate them to one or more control centers called sinks or base stations. The end-user typically accesses the data at the sinks. Some of the characteristics of a typical sensor node are limited computing capability, memory capacity and battery charge. Each sensor node operates with a limited transmission range, which is the distance until which the signals emanating from the node propagate and are received with appreciable signal strength. Wireless sensor networks operate with a limited bandwidth that is also shared among the nodes within a common transmission range. The sink is normally static and located far away from the sensor nodes. Hence, direct communication between the sensor nodes and the sink is expensive in terms of energy consumption and bandwidth usage. This forms the motivation to deploy data gathering algorithms at the sensor nodes to combine data into a small set of meaningful information, which is a representative of the network condition and can be transmitted to the sink, leading to significant energy and bandwidth savings. Throughout this paper, the terms 'data gathering', 'data fusion' and 'data aggregation' are used interchangeably. They mean the same.

In this paper, we consider a wireless sensor network wherein the sensor nodes periodically (in rounds) report the collected data to the sink. For each round of communication, each sensor node generates a data packet and wants the information to be transferred to the sink. If each sensor node directly transmits its data to the sink that is typically located far away from the network field, it would lead to significantly high energy consumption per round. To minimize energy consumption, it would be more efficient if the data packets from sensor nodes are gathered and aggregated with the packets of peer sensor nodes and only one aggregated data packet (of the same size as that of the individual data packets) is sent per round from the sensor network to the sink. With data aggregation, we can filter the uncorrelated noise in several signals and transmit to the sink a more accurate signal that is representative of the network condition.

Various data gathering algorithms have been proposed in the literature. Well-known among these are the LEACH (Low-Energy Adaptive Clustering Hierarchy) [1] and PEGASIS (Power-Efficient Gathering in Sensor Information Systems) [2] algorithms. Both of these algorithms operate through discrete rounds of data collection. LEACH operates in two phases: a set-up phase and a steady-state phase. In the set-up phase, the sensor nodes are grouped into clusters with the assignment of a cluster head for each cluster; in the steady-state phase, the sensor nodes transmit the collected data to their individual cluster heads and after gathering data from all of the sensor nodes in its cluster, a cluster head transmits the aggregated data to the sink. In PEGASIS, a chain of sensor nodes is formed using a greedy approach, starting from the node farthest to the sink. The nearest node to this node is added as the next node in the chain. This procedure is repeated until all the nodes are included in the chain. A node can be in the chain at only one position. During each round, a leader node is randomly selected and it is responsible for forwarding the aggregated data to the sink. After the sink selects the leader node and notifies it to the network, each node on both sides of the chain (with respect to the leader node), receives and transmits the aggregated data to the next node in the chain, until the data reaches the leader node. The original PEGASIS algorithm based on Time Division Multiple Access (referred to as PEGASIS-TDMA) resulted in huge delay as data moves across the complete chain of sensor nodes before being transmitted to the sink. PEGASIS has been later improved for CDMA (Code Division Multiple Access) systems [3] using a chain-based binary scheme to minimize the delay incurred and to reduce the energy*delay metric [4]. In PEGASIS-CDMA, a round comprises of $\log N$ levels where $N$ is the number of nodes in the network. For every round of data gathering, each node transmits data to a close neighbor in a given level of the hierarchy. Nodes that receive data at a given level are the only nodes that rise to the next level. At the top level, there will be only one node that will remain as the leader and it will transmit the aggregated data packet to the sink.

The distance-based chain formation heuristic of PEGASIS is prone to an increase in the physical distance between successive nodes as the chain progresses away from the starting node. This can lead to higher energy consumption per round. PEGASIS-CDMA consumes more energy per round compared to PEGASIS-TDMA, because the former requires nodes to communicate over long-distances because of the binary tree hierarchy. Nevertheless, the energy consumed per round for LEACH is significantly more than that consumed for PEGASIS-TDMA and PEGASIS-CDMA. If multiple cluster-heads are selected, even though the delay per round would be low because of reduced size of a cluster, several cluster-heads would be transferring data over long-distances to the sink. On the other hand, if few cluster-heads are selected, the sensor nodes may have to transmit over long-distances to reach the nearest cluster-head. Due to competition in each cluster, the delay might also increase. The above qualitative analysis of the energy consumption per round for the LEACH and PEGASIS algorithms suggests that the number of rounds sustained by the sensor network running these algorithms before the failure of the first sensor node (due to the exhaustion of battery charge) need not be maximum. There is a possibility of increasing the number of rounds before the first node failure by reducing the energy consumed per round and choosing only the nodes with a relatively higher energy level for data gathering from multiple sensor nodes and transmitting the aggregated data to an another sensor node or the sink. This observation formed the motivation of our work in this paper.

In this paper, we advocate the use of a maximal leaf nodes tree for energy-efficient data gathering in wireless sensor networks. A leaf node in a data gathering tree need to turn on itself for only to periodically sense and transmit data to its parent and the node can remain asleep for the rest of the time. On the other hand, an intermediate node would need to be turned on and stay in active and listening modes much longer than a leaf node because the intermediate node would need to receive data from all of its child nodes, aggregate with its own sensed data and transmit to its parent. This observation made in [5] motivates the need to consider a maximal leaf nodes spanning tree for data gathering in wireless sensor networks. In addition, the problem of determining a maximal leaf spanning tree has been proven to be polynomially equivalent to the problem of approximating a Minimum Connected Dominating Set (MCDS) [6]. The above two observations form the basis for our paper. A common approach (e.g., [7]) for approximating an MCDS is to consider inclusion of nodes that have a larger number of uncovered neighbors. We provide energy-awareness to this strategy by computing the weight of a node as the product of the number of uncovered neighbors and the available residual energy at the node. Such an energy-aware maximal leaf data gathering tree has not been proposed in the literature for wireless sensor networks.

Only a couple of works are available on the application of maximal leaf tree for efficient data gathering in wireless sensor networks. In [5][8], the authors compared maximal leaf tree (MLT) based data gathering and shortest path tree (SPT) based data gathering and observed that the MLT-based approach incurs an energy consumption and delay that is one-half of that incurred with the SPT-based approach in tree constructions, data transmissions and dynamic tree reconstructions. In [9], the authors advocate the use of a maximal leaf spanning tree for minimizing the number of forwarding nodes as part of an energy-efficient broadcast mechanism in wireless sensor networks. However, both of the above works attempt to only maximize the number of leaf nodes without paying attention to the energy-level of the intermediate nodes selected for data aggregation.

The rest of the paper is organized as follows: In Section 2, we describe the proposed algorithm to construct an Energy-aware Maximal Leaf Nodes Data Gathering (EMLN-DG) tree. Section 3 describes the algorithm to compute the delay per round for the EMLN-DG tree. Section 4 illustrates an example to construct the EMLN-DG tree and compute its delay. Section 5 describes the simulation environment and the performance results obtained comparing EMLN-DG with that of LEACH and PEGASIS (both TDMA and CDMA versions). Section 6 concludes the paper.

## 2 Description of the Algorithm to Construct the Energy-aware Maximal leaf Nodes Data Gathering (EMLN-DG) Tree

The algorithms to construct the EMLN-DG tree (pseudo code in Figure 1) and compute its delay (pseudo code in Figure 2) use the following variables associated with each node. The input to the algorithm is a snapshot of the underlying sensor network graph $G = (V, E)$ at a particular time instant during which we want to find the data gathering tree. Note that $V$ represents the set of vertices (nodes in the network) and $E$ represents the set of edges (links between the nodes in the network). There exists an edge between any two vertices in the graph if and only if the distance between the corresponding nodes in the network is less than or equal to the transmission range.

### 2.1 Variables associated with each Node

*Neighbors*($s$) – List of neighbors of node $s$ in graph $G$
*Uncovered-Neighbors*($u$) $\subseteq$ *Neighbors*($u$)
 – the list of neighbors of node $u$ that are not yet covered by the EMLN-DG tree
*Energy*($u$) – the residual energy (in Joules) available at node $u$
*Weight*($u$) – | *Uncovered-Neighbors*($u$) | * *Energy*($u$)
*Level*($u$) – the level of node $u$ in the EMLN-DG tree; the root node is at level 0.
*Child-Nodes*($u$) – the list of immediate child nodes (downstream nodes) of node $u$
*Predecessor-Node*($u$) – the predecessor node for node $u$ in the EMLN-DG tree

### 2.2 Algorithm to Construct the EMLN-DG Tree

The algorithm is executed for each round of data aggregation. The weight of a sensor node is the product of the number of uncovered neighbors of the node and the residual energy level available at the node. The sensor node that has the largest weight value is selected as the *rootNode* for the round. The algorithm uses the following four principal data structures to facilitate the computation of the DG tree:
(i) *Intermediate-Nodes-List* – the list of all the intermediate nodes, including the root node of the DG tree
(ii) *Leaf-Nodes-List* – the list of all the leaf nodes of the DG tree
(iii) *Covered-Nodes-list* – the list of all nodes that are spanned (i.e., covered) by the DG tree
(iv) *Priority-Queue* – a queue whose entries are tuples of the form (*Weight*($u$), $u$) and the entries are stored in the decreasing order of the node weights. A dequeue operation on this queue results in extracting the node with the minimum weight from the queue. The weight of a node is the product of the number of uncovered neighbors of a node and the residual energy available at the node, If two or more nodes have the same minimum weight, a node is randomly chosen and extracted from the queue.

The *Level* of the *rootNode* in the DG tree is set to 0. The maximum value for the *Level* of a node in the DG tree corresponds to the *Height* of the tree. To compute the delay associated with data gathering per round, we keep track of the list of nodes at each level of the tree through a data structure referred to as *Nodes-All-Levels*. As the DG tree is constructed by finding the intermediate nodes of the tree, the *Level* values of the nodes are determined and used to update the *Nodes-All-Levels* data structure.

Initially, the *Intermediate-Nodes-List*, *Leaf-Nodes-List* and *Covered-Nodes-List* are initialized to null set (i.e., empty). The *Priority-Queue* is populated with entries in the form of the tuple (*Weight*($u$), $u$), with the entries sorted in the decreasing order of the values of the node weights. The *rootNode* is the first node to be dequeued from the *Priority-Queue* and added to the *Covered-Nodes-List* and the *Intermediate-Nodes-List*. Once a node is added to the *Intermediate-Nodes-List*, all its neighboring nodes are considered covered, and added to the *Covered-Nodes-List* if not previously covered. If a node is added to the *Covered-Nodes-List* as a result of being newly covered by its predecessor node, then the node is also added to the *Leaf-Nodes-List*. Later, if a covered node is selected as an intermediate node, it is removed from the *Leaf-Nodes-List* and added to the *Intermediate-Nodes-List*.

---

**Input:** Snapshot of the Network Graph $G = (V, E)$, where $V$ is the set of vertices and $E$ is the set of edges

**Auxiliary Variables and Functions:**
*Intermediate-Nodes-List*, *Leaf-Nodes-List*, *Covered-Nodes-List*, Priority-Queue
*rootNode* – the first node to be added to the *Intermediate-Nodes-List*
*intermediateNode* – the next node that could be added to the *Intermediate-Nodes-List*
*Nodes-At-Level*($i$) – the list of nodes at a particular level $i$ in the DG tree
*Nodes-All-Levels* – the list of tuples ($i$, *Nodes-At-Level*($i$))

**Output:** *Intermediate-Nodes-List*, *Leaf-Nodes-List*, *Nodes-All-Levels*, *Height-DG-Tree*, *rootNode*
                                            // if the underlying Network Graph $G$ is connected
       NULL // if the underlying Network Graph $G$ is not connected

**Initialization:**
 *Intermediate-Nodes-List* = Φ; *Leaf-Nodes-List* = Φ; *Covered-Nodes-List* = Φ;
 $\forall u$, *Uncovered-Neighbors*($u$) = *Neighbors*($u$); *Nodes-All-Levels* = Φ; *Height-DG-Tree* = 0
 **for** every vertex $u \in V$ **do**   // can be implemented in O($|V|*\log|V|$) time
    Insert the tuple (*Weight*($u$), $u$) to the *Priority-Queue* at the appropriate location in the queue
              // It takes O($\log|V|$) time to insert an entry at the appropriate location
    *Level*($u$) = -1
    *Child-Nodes*($u$) = Φ
    *Predecessor-Node*($u$) = NULL
 **end for**

**Begin** EMLN-DG Construction

*rootNode* = Dequeue(*Priority-Queue*)
*Level*(*rootNode*) = 0
*Covered-Nodes-List* = *Covered-Nodes-List* U {*rootNode*}
*Intermediate-Nodes-List* = *Intermediate-Nodes-List* U {*rootNode*}
*Nodes-At-Level*(0) = *Nodes-At-Level*(0) U {*rootNode*}
*Nodes-All-Levels* = *Nodes-All-Levels*(0, *Nodes-At-Level*(0))

**for** every vertex *v* ∈ *Neighbors*(*rootNode*) **do**
   *Covered-Nodes-List* = *Covered-Nodes-List* U {*v*}
   *Leaf-Nodes-List* = *Leaf-Nodes-List* U {*v*}
   *Level*(*v*) = 1
   *Child-Nodes*(*rootNode*) = *Child-Nodes*(*rootNode*) U {*v*}
   *Predecessor-Node*(*v*) = *rootNode*

   **if** ( *Height-DG-Tree* < *Level*(*v*) ) **then**
      *Height-DG-Tree* = *Level*(*v*)
   **end if**

**end for**

**while** ( |*Covered-Nodes-List*| < |*V*| ) **do** // takes O(|*V*|*(|*V*|+|*E*|)) time with O(|*V*|+|*E*|) per iteration
   **for** every vertex *u* ∈ *V* **do** // takes O(|*V*|+|*E*|) time
     **for** every vertex *v* ∈ *Neighbors*(*u*) **do**
       **if** *v* ∈ *Covered-Nodes-List* **then**
         *Uncovered-Neighbors*(*u*) = *Uncovered-Neighbors*(*u*) – {*v*}
       **end if**
     **end for**
   **end for**

   *Priority-Queue* = Φ; // Empty the *Priority-Queue* – can be done in O(|*V*|) time

   **for** every vertex *u* ∈ *V* and *u* ∉ *Intermediate-Nodes-List* **do** // can be done in O(|*V*|*log|*V*|) time
     *Weight*(*u*) = |*Uncovered-Neighbors*(*u*)| * *Energy*(*u*)
     Insert the tuple (*Weight*(*u*), *u*) to *Priority-Queue* at the appropriate location in the queue
   **end for**

   **if** (*Priority-Queue* = Φ) **then**
     **return** NULL; // the underlying network graph *G* is not connected
   **end if**

   *intermediateNode* = Dequeue(*Priority-Queue*)

   *Intermediate-Nodes-List* = *Intermediate-Nodes-List* U {*intermediateNode*}
   *Leaf-Nodes-List* = *Leaf-Nodes-List* – {*intermediateNode*}

   **for** every vertex *v* ∈ *Neighbors*(*intermediateNode*) **do**
      **if** (*v* ∉ *Covered-Nodes-List*) **then**

```
            Predecessor-Node(v) = intermediateNode
            Level(v) = Level(intermediateNode) + 1
            Covered-Nodes-List = Covered-Nodes-List U {v}
            Leaf-Nodes-List = Leaf-Nodes-List U {v}
            Child-Nodes(intermediateNode) = Child-Nodes(intermediateNode) U {v}
            if ( Height-DG-Tree < Level(v) ) then
                Height-DG-Tree = Level(v)
            end if
         end if
      end for

      Nodes-At-Level(Level(intermediateNode)) = Nodes-At-Level(Level(intermediateNode)) U
                                        {intermediateNode}

end while

return Intermediate-Nodes-List, Leaf-Nodes-List, Nodes-All-Levels, Height-DG-Tree, rootNode

End EMLN-DG Construction
```

**Figure 1:** Algorithm to Construct the EMLN-DG Tree

After we populate the *Covered-Nodes-List* by exploring the neighbors of the *rootNode*, we use this list as the base and proceed to cover the rest of the nodes in the network by executing a loop – an iteration of the loop has the following sequence of steps:

Step 1 – Re-compute the set of *Uncovered-Neighbors* for each node: For each vertex *u* in the graph, if a neighbor *v* has been newly added to the *Covered-Nodes-List* (i.e., the neighbor *v* was uncovered before the previous iteration), then vertex *v* is removed from *u*'s set of *Uncovered-Neighbors*.

Step 2 – Re-compute the *Priority-Queue*: The *Priority-Queue* is reset to null set (i.e., empty). The *Weight* of the covered nodes *u* (i.e., nodes in the *Covered-Nodes-List*) that are not in the *Intermediate-Nodes-List* is computed and the tuple (*Weight(u)*, *u*) is stored at the appropriate location in the queue. Note that in order for a node to be considered for inclusion in the *Priority-Queue* and considered a candidate for being selected as an intermediate node, the node has to be already covered (i.e., be part of the *Covered-Nodes-List*) by another intermediate node. This is essential to maintain the connectivity of the network. If the *Priority-Queue* continues to be empty (i.e., none of the covered nodes have uncovered neighbors), then we stop the execution of the algorithm and return NULL – i.e., the underlying sensor network is not connected.

Step 3 – Add a node to the *Intermediate-Nodes-List*: We dequeue the *Priority-Queue* and add the extracted node to the *Intermediate-Nodes-List* and remove it from the *Leaf-Nodes-List*. Every uncovered neighbor of the newly determined intermediate node is added to the *Covered-Nodes-List* and the *Leaf-Nodes-List*, added as a child node for the intermediate node. The *Level* of a newly covered node is one more than the *Level* of its immediate predecessor intermediate node through which the former was covered.

If the algorithm runs until all the nodes in the network are covered, it returns the following five data structures/variables that are used to compute the delay per round associated with the DG tree: *Intermediate-Nodes-List*, *Leaf-Nodes-List*, *Nodes-All-Levels*, *Height-DG-Tree* and *rootNode*. The time complexity of the algorithm is $O(|V|*(|V|+|E|))$, as it takes $O(|V|+|E|)$ time per iteration. At the beginning of each iteration (i.e. in Step 1), we have to re-compute the number of uncovered neighbors for every node in the network. There are $|V|$ nodes in the network and we have to process each of the $|E|$ edges twice, once for each vertex on which the edge is incident.

## 3  Algorithm to Compute the Delay per Round for the EMLN-DG Tree

The delay at a node indicates the number of time slots it takes for the root node to receive the aggregated data from all of its immediate child nodes. The delay associated with each of the leaf nodes is 0. We assign one time slot per child node to transfer data to its immediate predecessor intermediate node. We start processing the intermediate nodes from the bottom of the DG tree. Note that the intermediate nodes at a particular level in the DG tree are independent of each other and their delay can be computed in parallel. Also, a node is the child node for only one intermediate node in the DG tree.

For each intermediate node $u$ at a particular level, we prepare a sorted list of the delay associated with each of its immediate child nodes. The delay associated with the intermediate node is computed through a temporary running variable, *Temp-Delay* (initialized to zero), as we explore the delay associated with each of the child nodes in the sorted list. For every child node $v$ in the sorted list of the delay, *Temp-Delay* is set to the maximum of *Temp-Delay* + 1 and *Delay*($v$) + 1, as we assume it takes one time slot for a child node to transfer its aggregated data to the immediate predecessor intermediate node. The delay associated with the intermediate node $u$, *Delay*($u$), is the final value of *Temp-Delay* after we run through the sorted list of the delays associated with the *Child-Nodes*($u$). The above procedure is repeated for all intermediates nodes, from levels one less than the *Height* of the tree all the way to zero (i.e. the root node). The delay per round for the DG tree is the delay associated with the *rootNode*. The delay per round on a $|V|$-node DG tree can be computed in $O(|V|*\log|V|)$ time as there are only $|V|$-1 edges in the tree and $O(\log|V|)$ is the time it takes to update additions to a sorted list that is implemented as a heap.

---

**Input:** *Intermediate-Nodes-List*, *Leaf-Nodes-List*, *Nodes-All-Levels*, *rootNode*
**Output:** *Delay-per-Round*
**Auxiliary Variables:**
   *Delay*($u$) // Number of time slots it takes for the aggregated data to reach node $u$
   *Node-Level*    // level of a particular node in the DG tree
   *Nodes-At-Level*    // the list of nodes at a particular level
   *Sorted-Delay-Child-Nodes* // sorted list (in the increasing order) of the delay of the child nodes
   *Temp-Delay* // temporary variable to process the delay at a node
**Initialization:** *Delay-per-Round* = 0

**Begin** Computation-Delay-EMLN-DG-Tree
   **for** (every vertex $v \in$ *Leaf-Nodes-List*) **do**
      *Delay*($v$) = 0
   **end for**
   **for** (*Node-Level* = *Height-DG-Tree*-1 to 0) **do** // can be done in $O(|V|*\log|V|)$ time
      *Nodes-At-Level* = *Nodes-All-Levels*(*Node-Level*)
      **for** (every vertex $u \in$ *Nodes-At-Level*) **do**
         *Temp-Delay* = 0
         *Sorted-Delay-Child-Nodes* = Φ
         **for** (every vertex $v \in$ *Child-Nodes*($u$)) **do**
            Insert the tuple {$v$, *Delay*($v$)} at an appropriate entry in *Sorted-Delay-Child-Nodes*
         **end for**
         **for** (every tuple {$v$, *Delay*($v$)} in the *Sorted-Delay-Child-Nodes* list) **do**
            *Temp-Delay* = Maximum (*Temp-Delay* + 1, *Delay*($v$) + 1)
         **end for**
         *Delay*($u$) = *Temp-Delay*
      **end for**

**end for**

**return** *Delay(rootNode)*

**End** Computation-Delay-EMLN-DG-Tree

**Figure 2:** Algorithm to Compute the Delay per Round for the EMLN-DG Tree

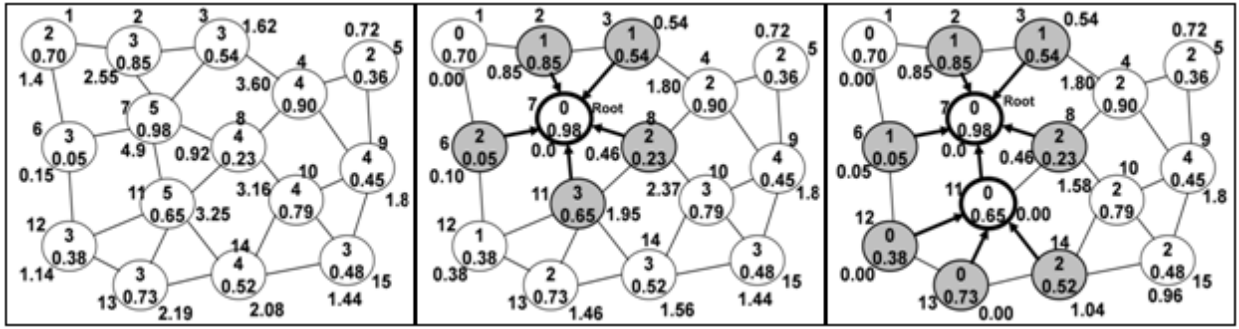

Figure 3.1: Initial Network    Figure 3.2: Iteration # 1    Figure 3.3: Iteration # 2

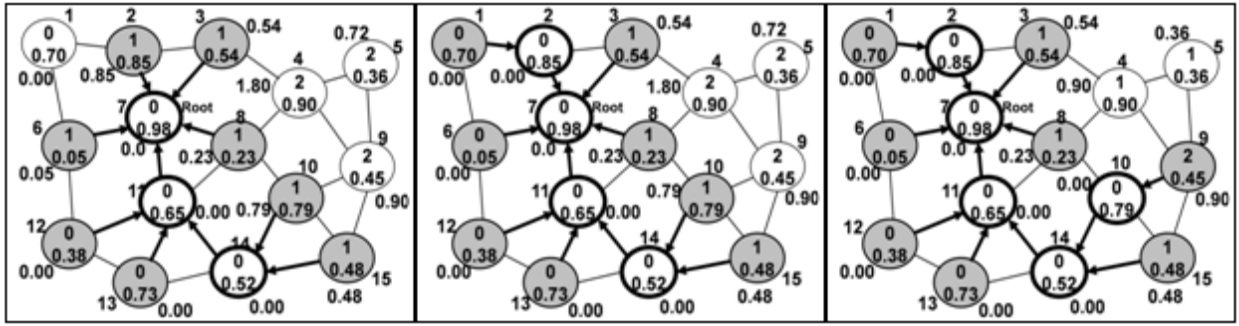

Figure 3.4: Iteration # 3    Figure 3.5: Iteration # 4    Figure 3.6: Iteration # 5

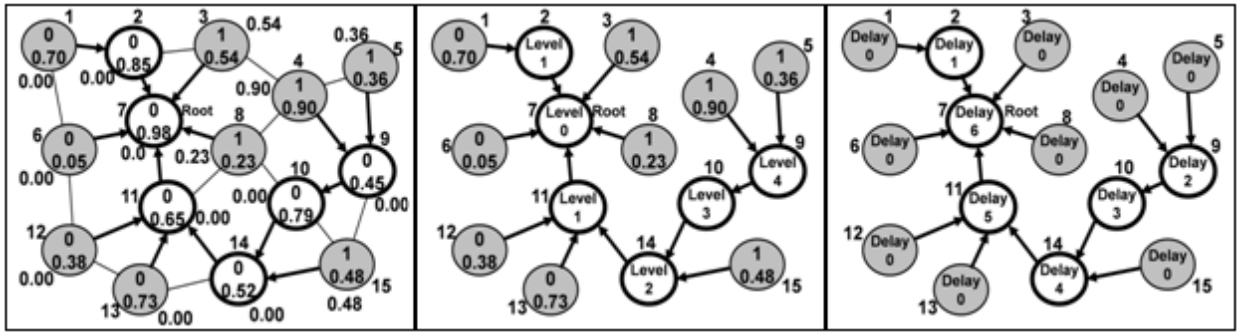

Figure 3.7: Iteration # 6    Figure 3.8: Final EMLN-DG Tree    Figure 3.9: Delay Computation

**Figure 3:** Example Illustrating the Construction of the EMLN-DG Tree and the Computation of its Delay

## 4   Example to Illustrate the Construction of EMLN-DG Algorithm

Figure 3 illustrates an example to demonstrate the working of the EMLN-DG algorithm. In Figures 3.1 – 3.9, each circle represents a node. The integer outside the circle represents the node ID and the integer inside the circle represents the number of uncovered neighbors of the corresponding node. The real-number inside the circle represents the residual energy (in Joules) currently available at the node and the real-number outside the circle represents the weight (product of the residual energy and the number of uncovered neighbors) for the particular node. The intermediate nodes that are part of the EMLN-DG tree have their circles bold. We shade the circles of leaf nodes that are covered by the intermediate nodes of the EMLN-DG tree. The circles of nodes that are not yet part of the EMLN-DG tree (i.e., nodes that are neither intermediate nodes nor leaf nodes) are neither shaded nor made bold.

On the 15-node example illustrated in Figure 3, we observe that it takes 6 iterations to compute the final EMLN-DG tree. The EMLN-DG tree has only 6 intermediate nodes, including the root node. The remaining 9 nodes (i.e. 60% of the nodes) are leaf nodes. The height of the tree is 5 (nodes 4 and 5 are at level 5) and the delay at the root node (i.e. number of time slots it takes for the root node 7 to receive the aggregated data from all its immediate downstream child nodes) is 6 time slots.

## 5   Simulations

We evaluated the performance of the EMLN-DG algorithm vis-à-vis LEACH, PEGASIS-TDMA and PEGASIS-CDMA in a discrete-event simulator developed by us in Java. This simulator has been successfully used to develop and study data gathering algorithms in some of our recent studies [10][11][12]. The network dimension is 100m x 100m. The number of nodes in the network is 100 and they are uniformly and randomly distributed throughout the network. The sink node is located away from the sensor field, at location (50, 300). In the case of EMLN-DG, the transmission range per sensor node adopted for forming the network graph in the algorithm is varied from 15m to 50m. There exists an edge between any two vertices in the graph if the distance between the corresponding nodes in the network is less than or equal to the transmission range. Sensor nodes running the LEACH, PEGASIS-TDMA and PEGASIS-CDMA algorithms are assumed to be able to conduct transmission power control (i.e. vary their transmission range) depending on the distance to the receiver node.

### 5.1 Assumptions for EMLN-DG

We assume all the sensor nodes are CDMA (Code Division Multiple Access) enabled so that we can achieve parallel communication between any pair of sensor nodes as and when desired. Such an assumption has also been made in other well-known data aggregation algorithms such as LEACH and PEGASIS. For data aggregation in the EMLN-DG tree, we assume that every intermediate upstream node uses a particular CDMA code to communicate with all its immediate downstream child nodes. The upstream node broadcasts a time schedule for data transmission to all its immediate downstream nodes. A downstream node sends its data to the upstream node according to the slots provided in the time schedule. Note that such TDMA (Time Division Multiple Access) – based communication between every upstream node and its downstream nodes can occur in parallel using the unique CDMA codes chosen by each of the upstream nodes.

### 5.2 Energy Consumption Model

We will use the following first order radio model [13] that has been also used to model energy consumption in the previous work [1][2][4]. According to this model, the energy expended by a radio to run the transmitter or receiver circuitry is $E_{elec}$ = 50 nJ/bit and $\in_{amp}$ = 100 pJ/bit/m$^2$ for the transmitter amplifier. The radios are turned off when a node wants to avoid receiving unintended transmissions. An $r^2$ energy loss model is used to compute the transmission costs. The energy lost in transmitting a *k*-bit message over a distance *d* is given by: $E_{TX}(k, d) = E_{elec} * k + \epsilon_{amp} * k * d^2$. The energy lost in receiving a *k*-bit message is $E_{RX}(k) = E_{elec} * k$.

The energy lost per round is the sum of the energy lost at all the nodes for the transmission, reception and fusion of the data. The leaf nodes in the EMLN-DG tree do not lose energy to receive or fuse the data, but lose energy to transmit data to their upstream node. Every intermediate node, including the Root node, loses energy to receive and aggregate data from each of its immediate downstream nodes and to forward the aggregated data to its upstream node in the tree. Note that the sink can be considered as the upstream node for the Root node.

### 5.3 Performance Metrics

The performance metrics considered are: (i) Energy lost per round, (ii) Delay (in terms of the number of time units) per round of data aggregation and transmission to the sink, (iii) Network lifetime, measured as the number of rounds the network sustains before the first sensor node dies due to exhaustion of battery charge and (iv) energy*delay value per round. For the EMLN-DG tree, we also evaluate the impact of the transmission range on the (i) Probability of connectivity of the tree, (ii) Energy lost per round, (iii) Delay per round, (iv) Energy*Delay per round, (v) Number of leaf nodes per tree and (vi) Network lifetime, measured as the round of first node failure. The results reported in Figure 4 for EMLN-DG are obtained for 1000 trials of the algorithm for each value of the transmission range per node. The results reported in Figure 5 for EMLN-DG correspond to a transmission range of 25m as the algorithm appears to give the best performance with respect to several metrics at this transmission range value. The results reported in Figure 5 for LEACH, PEGASIS-TDMA and PEGASIS-CDMA versions do not assume a particular transmission range per node as these three algorithms assume that the nodes can do transmission power control as and when needed. For each trial, the initial energy supplied to every sensor node is 1J.

### 5.4 Impact of Sensor Transmission Range on the Performance of EMLN-DG

Figure 4.1 illustrates that the probability of network connectivity reaches 0.994 (99.4%) when the sensor transmission range is 25m. Lower values of transmission range result in poor network connectivity and as the transmission range increases, the connectivity increases. After the transmission range value reaches 25m, the probability of network connectivity does not change much, increases only from 0.994 to 0.9998 as we increase the transmission range from 25m to 35m. The height of the DG tree decreases with increase in the sensor transmission range and this can be attributed to the corresponding increase in the number of leaf nodes (and a decrease in the number of intermediate nodes). As the transmission range increases, the number of nodes that can be covered by the inclusion of a node as an intermediate node in the DG tree increases. Hence, less number of intermediate nodes needed to be added in the DG tree. On a 100-node network, we observe that only less than 35% of the nodes serve as intermediate nodes of the EMLN-DG tree at a very low transmission range per node value of 15m and as the transmission range per node increases, the percentage of nodes serving as intermediate nodes significantly decreases. At a transmission range of 50m (i.e., half of the one-side dimension of the network), less than 4% of the nodes serve as intermediate nodes of the tree.

For low and moderate values of transmission range, more intermediate nodes are added to the DG tree and this leads to less number of downstream child nodes per upstream node in the DG tree. On the other hand, for higher transmission range values, less number of intermediate nodes are included in the DG tree and each upstream intermediate node in the DG tree has several downstream child nodes. There can be simultaneous transmissions across the different upstream node – downstream node sets, with each upstream node assigning a unique CDMA code to its set of downstream nodes. In order to avoid too much of interference with simultaneous transmissions, we assume that all the downstream nodes of an upstream node send their data to the upstream node using a single CDMA code according to a time schedule and only one transmission is allowed per time unit (this is also the approach adopted for communication within a cluster in LEACH). An upstream node has thus to wait to receive data from all its downstream nodes before propagating the data further. The more the number of downstream nodes, more the delay incurred at an upstream node to collect, aggregate and further transfer the data.

The energy lost per round increases with increase in the sensor transmission range as more energy is expended with increase in the distance. But, the increase in the energy lost per round is not proportional to the increase in the sensor transmission range. As the transmission range is doubled (to 30m) and tripled (to 45m) from 15m, the energy lost per round is at most 18% and 49% respectively. This can be attributed to the reduction in the number of leaf nodes in the DG tree with increase in the sensor transmission range and not all sensors need to do data aggregation, all the time.

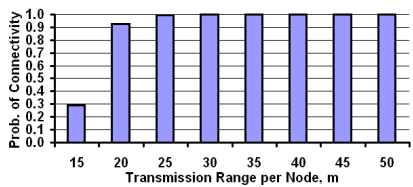 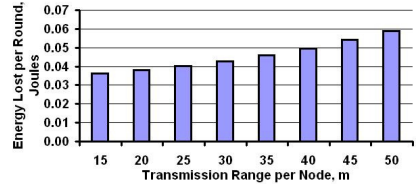 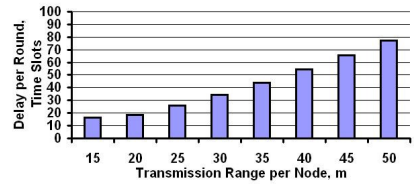

**Figure 4.1:** Connectivity   **Figure 4.2:** Energy Lost per Round   **Figure 4.3:** Delay per Round

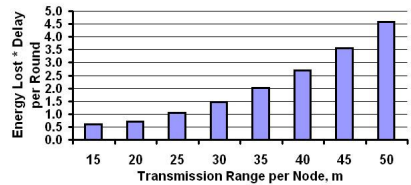 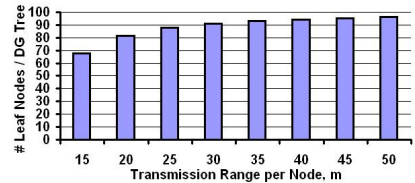 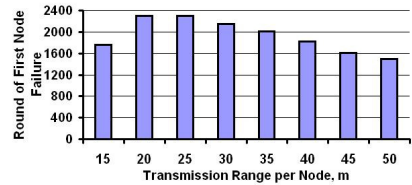

**Figure 4.4:** Energy*Delay per Round   **Figure 4.5:** # Leaf Nodes per DG Tree   **Figure 4.6:** Round of First Node Failure

**Figure 4:** Performance of the Energy-aware Maximal leaf Nodes Data Gathering (EMLN-DG) Tree

The network lifetime is the maximum when the transmission range per node is at 20m and 25m. As we increase the sensor transmission range from 25m to 35m and 50m, the network lifetime (in terms of the number of rounds) decreases, by about 12% and 35% respectively. This can be attributed to a slight to moderate increase in the energy lost per round. However, when we compute the energy*delay values, it remains low for smaller and moderate transmission range values and increases rapidly as the transmission range gets high.

Considering the performance of EMLN-DG with respect to all the performance metrics, we conclude that the optimal transmission range per sensor to achieve higher connectivity and lower energy*delay value for a 100-node network of dimensions 100m x 100m is 25m. This translates to a neighborhood size of $\frac{\frac{22}{7}*(25m)^2*100 nodes}{100m*100m} \approx 20$ nodes. At this transmission range value of 25m, less than 15% of the nodes in the network serve as intermediate nodes and the rest of the nodes serve as leaf nodes of the tree. The performance comparison studies reported for EMLN-DG in Section 5.5 are based on this transmission range value of 25m per node.

**5.5 Performance Comparison Study with LEACH and PEGASIS**

EMLN-DG incurs the largest value for the network lifetime (the number of rounds the network runs before the first node failure occurs due to exhaustion of battery charge). This could be attributed to the balancing effect due to the consideration of the energy levels of the nodes as well as their number of uncovered neighbors before deciding whether a node would serve as an intermediate node or leaf node. Most of the intermediate nodes added to the EMLN-DG tree have relatively higher energy than the leaf nodes of the tree. The leaf nodes spend relatively less energy for data transfer per round whereas the intermediate nodes spend more energy for data reception, aggregation and transmission per round. Nevertheless, with the EMLN-DG tree being refreshed and determined for every round, we achieve a very high energy balance in the network. This leads to a significantly high network lifetime and is much larger compared to those obtained with the contemporary protocols like LEACH and PEGASIS. As future work, we will study the performance of EMLN-DG when the DG tree is not computed for every round, but updated only periodically, for every $n$ ($n > 1$) number of rounds.

With LEACH, even though the clusters are updated for every round of data aggregation, the cluster heads have to transmit the aggregated data over a longer distance to the sink node. In the case of PEGASIS and EMLN-DG, only the leader node/root node is responsible for transferring the aggregated data to the sink node. The energy lost per round for EMLN-DG is 62% less than that incurred for LEACH and the network lifetime for EMLN-DG is about 235% greater than that obtained for LEACH.

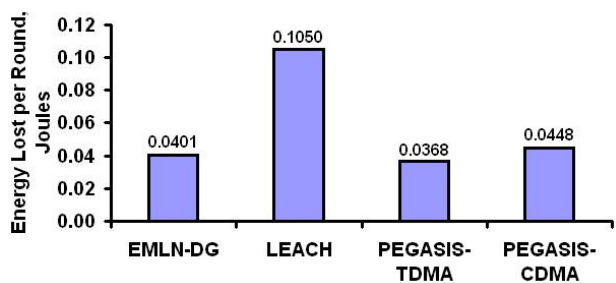
**Figure 5.1:** Energy Lost per Round, Joules

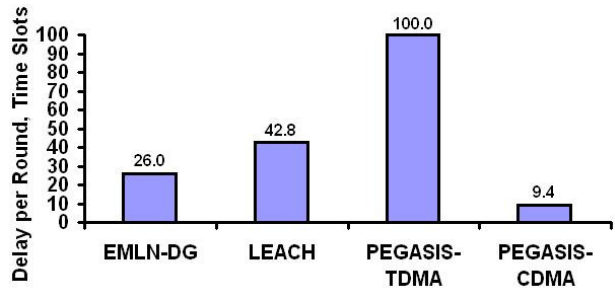
**Figure 5.2:** Delay per Round, Time Slots

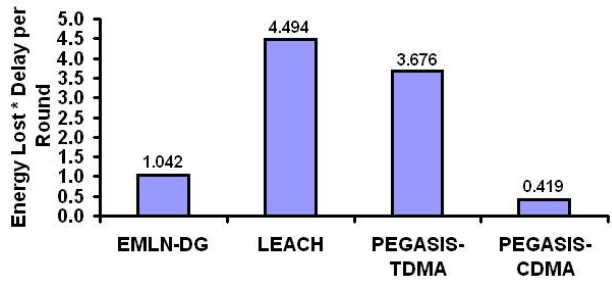
**Figure 5.3:** Energy Lost * Delay per Round

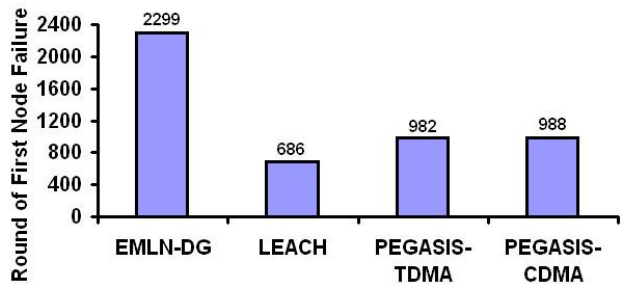
**Figure 5.4:** Round of First Node Failure

**Figure 5:** Performance Comparison of EMLN-DG, LEACH, PEGASIS (TDMA and CDMA Systems)

The reasons that can be attributed to the relatively lower lifetime for the PEGASIS – TDMA and CDMA versions are that the distance-based chain is formed only once and is being used for all the rounds and also, the average distance between successive nodes in the second half of the chain has been found to be far more than the average distance between successive nodes in the first half of the chain. This is due to suboptimal solution obtained with the greedy distance-based chain formation heuristic. The energy consumption per round for PEGASIS-TDMA is relatively lower (actually the lowest of all the four algorithms considered) than that of PEGASIS-CDMA because, in the latter case, nodes that are far away from each other are more likely required to communicate because of the binary-tree approach. The energy consumption per round for EMLN-DG is about 10% more than that incurred for PEGASIS-TDMA and is about 10% less than that incurred for PEGASIS-CDMA. Nevertheless, because of the huge energy imbalance created due to the use of a static and suboptimal greedy distance-based chain (without any consideration of the energy levels of the nodes in the chain), the network lifetime per round for PEGASIS is far lower than that of EMLN-DG. We observe that the network lifetime per round for EMLN-DG is about 135% more than that obtained for the two PEGASIS versions.

EMLN-DG incurs a larger delay than that of PEGASIS-CDMA and this can be attributed to the use of time-division multiplexing during the aggregation of the data packets at an upstream node from its immediate downstream nodes. PEGASIS-CDMA assumes that at a given level in the binary tree hierarchy, any two nodes in the chain can communicate simultaneously with each other using a unique CDMA code. Even though CDMA codes help to avoid mixing up of the data signals, there can be still some interference among the signals if all the simultaneous communications are targeted to a single node. Hence, we assume the use of time-division multiplexing in EMLN-DG for data aggregation at every upstream node. All the downstream nodes of an upstream node forward data to the upstream node according to a time schedule, with only one transmission allowed per time slot. Data aggregation at different upstream nodes can however occur in parallel, using different CDMA codes. Also, an intermediate node in the EMLN-DG tree cannot forward the aggregated data to its upstream node until it receives data from each of its immediate downstream nodes. The delay per round of communication for EMLN-DG is about 2.75 times to that incurred with PEGASIS–CDMA. However, the delay incurred for EMLN-DG is still lower than that incurred for PEGASIS–TDMA and LEACH. The delay incurred for PEGASIS–TDMA and LEACH is respectively about 3.85 and 1.65 times more than that incurred for EMLN-DG.

The energy*delay value is the least for the binary-tree based PEGASIS-CDMA. As discussed before, PEGASIS-CDMA incurs the lowest delay per round because of the assumption of simultaneous communication between any pair of nodes all the time. EMLN-DG incurs a larger energy*delay value, which is 2.5 times more than that incurred with PEGASIS-CDMA. On the other hand, the energy*delay values for PEGASIS-TDMA and LEACH is respectively 3.5 and 4.3 times more than that incurred for EMLN-DG.

Comparing the performance of EMLN-DG with that of direct transmission (where each sensor node directly transmits the data to the sink), we observe that (i) the energy lost per round in direct transmission is at least 30 times more than that incurred with EMLN-DG, (ii) the delay per round of data communication is about 4 times more than that incurred with EMLN-DG, (iii) the network lifetime incurred with direct transmission is less than $1/40^{th}$ of that incurred with EMLN-DG and (iv) the energy*delay value incurred with direct transmission is at least 120 times more than that incurred with EMLN-DG.

## 6   Conclusions

The high-level contribution of this paper is the development of an Energy-aware Maximal Leaf Nodes Data Gathering (EMLN-DG) algorithm that is aimed at simultaneously reducing the energy lost per round of data gathering as well as maximizing the number of rounds of communication before the first failure of a sensor node due to the exhaustion of battery charge. The overall run-time complexity of the ECDS-DG algorithm is $O(|V|*(|V|+|E|))$ where $|E|$ is the number of edges and $|V|$ is the number of vertices in the underlying sensor network graph. Performance simulation studies on EMLN-DG illustrate that when running the algorithm at a transmission range value that is one-fourth of the one-side dimension of the network employed, we observe that less than 15% of the nodes serve as intermediate nodes of the tree and rest of the nodes serve as leaf nodes. Even while operating the network at very low transmission range values per node, we observe that less than 35% of the nodes serve as intermediate nodes of the tree. Based on the simulation comparison studies of EMLN-DG with that of the well-known data gathering algorithms such as LEACH and PEGASIS (both TDMA and CDMA versions), we can conclude that EMLN-DG is a better data gathering algorithm as it incurs the maximum value for the network lifetime that is significantly high (as large as 200%-300% more) compared to those incurred by the other algorithms. At the same time, EMLN-DG also incurs a relatively lower, or if not comparable, energy loss, delay and energy*delay per round of data gathering vis-à-vis LEACH and PEGASIS.